\documentclass{iopart}
\usepackage[utf8]{inputenc}
\usepackage{graphicx}
\usepackage{xcolor}

\begin{document}
\title{Particle yields in pp interactions at $\sqrt{s}=17.3$ GeV 
interpreted in the statistical hadronization model}
\author{Tomasz Matulewicz and Krzysztof Piasecki\\ 
Institute of Experimental Physics, Faculty of Physics, University of Warsaw}
\date{\today}

\begin{abstract}

The unified set of yields of particles produced in proton-proton collisions at $\sqrt{s}$ = 17.3 GeV (laboratory beam momentum 158 GeV/c) is evaluated, combining the experimental results of the NA49 and NA61/SHINE collaborations at the CERN SPS.
With the statistical hadronization code Thermal-Fist 
we confirm the unacceptably high value of $\chi^2$, 
both in the canonical and grand canonical - strangeness canonical approach, 
and the common volume for all the hadrons.
The use of the energy-dependent width of the
Breit-Wigner parametrization for the mass distributions of unstable particles improves the quality of the description of particle yields only slightly.
We confirm the observation that exclusion of the $\phi$ meson yield makes the fit result acceptable.
The complete experimental data set of particle yields can be reasonably fitted if the canonical volumes of hadrons without and with open strangeness are allowed to vary independently.
The canonical volume of strangeness was found larger than that for non-strange hadrons, which is compatible with the femtoscopy measurements of p+p system at  $\sqrt{s} = $ 27.4 MeV and 900 MeV.
The model with the best-fit parameters allows to predict the yields of several not yet measured particles emitted from p+p at $\sqrt{s}$ = 17.3 GeV.

\end{abstract}

\maketitle

%---------------------------------------------
%-----------  introduction -------------------
%---------------------------------------------

\section{Introduction}

Statistical hadronization model is found to be very effective in the understanding 
of particle yields in the relativistic collisions of heavy nuclei 
\cite{Andronic2018, Cleymans2006}.
The applicability of the statistical model to small systems like proton-proton 
or even electron-positron collisions \cite{eehadronization, Becattini08} 
has also been studied. 
Proton-proton (pp) interactions were interpreted in statistical hadronization
models at energies available at SPS, RHIC and LHC 
\cite{Becattini1997,Kraus2010, Floris2014, Castorina2016, Begun2018}.
The extracted thermal model parameters show little energy dependence 
(except the baryochemical potential $\mu_B$): 
the chemical freeze-out temperature, $T$, remains close to 165 MeV, 
and the strangeness undersaturation factor $\gamma_S$ stays roughly within 
0.5 to 0.9 range \cite{Castorina2016}.

A decade ago the statistical hadronization in the pp system at $\sqrt{s} = 17.3$ GeV 
was analysed \cite{Kraus2010} using the data obtained by the NA49 Collaboration.
This research was stimulated by the discrepancies in thermal model parameters 
obtained in \cite{Kraus2009} and \cite{Becattini2006} for this collision system.
Basing on the available data set, the authors of \cite{Kraus2010} observed, 
that the results of thermal model were sensitive to the differences 
in data selection scheme (several sets were used). 
It was found, that the inclusion of the hidden-strangeness $\phi$ meson 
lowered down $\gamma_S$ and increased $T$.
However, if both $\phi$ meson and hyperon yields were included,
the reasonable description of the full data set was not possible. 

Since the publication of the analysis \cite{Kraus2010}, the NA49 collaboration 
concluded studies on the emission of baryons and strange mesons 
\cite{NA49Anticic2010b, NA49Anticic2010a, NA49Anticic2011} 
(in a few cases they slightly differed from the values in table I of \cite{Kraus2010}).
The successor of NA49, the NA61/SHINE Collaboration 
measured total yields of several hadrons emitted from the same colliding system 
\cite{NA61hminus, NA61lambda, NA61many, NA61phi, NA61Kstar, NA61Ksi}. 
The results of NA49 and NA61/SHINE Collaborations were used
(as separate data sets) in an extensive hadronization model calculations \cite{Begun2018}. 
A comparison of the results of this analysis to the experimental data provided 
high values of $\chi^2/NDF$, and for the canonical ensemble version of the model 
the extracted $\gamma_s$ undersaturation parameters were incompatible
between the data sets. 
However, if the yield of $\phi$ meson was removed from each set,
the fitted predictions of the same model were characterized by $\chi^2/NDF$ 
considerably lower (by the factor of 2.5), and were mutually consistent.
However, at that stage the published yield of $\phi$ mesons was only preliminary.
It should be also mentioned, that for the nucleus-nucleus (AA) collisions
at energies from SIS18 \cite{Agakishiev16} through AGS and SPS \cite{Becattini2006} 
up to LHC \cite{Andronic2018}, the statistical hadronization model fits well 
to the experimental data where the $\phi$ meson yields are included. 
On the other hand, the authors of paper \cite{Begun2018} studied the dynamical
fluctuations in the system of collided proton pairs. 
They found that it does not behave like a single statistical system 
with all charged conserved exactly in each event.

The extraction of yields of $\phi$ \cite{NA61phi} 
(the final value is close to the preliminary one used in \cite{Begun2018})
and $K^\star (892)^0$ \cite{NA61Kstar} mesons, as well as $\Xi$ and
$\bar{\Xi}$ baryons \cite{NA61Ksi} were concluded by the NA61/SHINE collaboration.
The measurements of NA61/SHINE collaboration supplement some of the NA49 results, 
providing also detailed evaluation of systematic uncertainties.
The collected results (exclusively from the published papers) 
is the basis of an evaluation of a combined data set of hadron yields, 
proposed in this paper.

For the model calculations we used Thermal-FIST, the recently available computer code for 
the statistical analysis of hadron production \cite{ThermalFIST}.
This code optionally offers for the resonances the energy-dependent widths ("eBW") 
of the Breit-Wigner mass distributions. This two-body decay-motivated recipe,
aimed to better describe the threshold effects of decays of broad resonances,
was first implemented in \cite{SHARE}, but is absent in the widely 
used THERMUS code \cite{THERMUS}. 
The recent study of the effect of the eBW approach on the thermal parameters
extracted for yields from Pb+Pb collisions at LHC energies shows, that it improves 
the $\chi^2$ values of fits, and slightly increases the temperature \cite{Vovchenko18}.
According to figure 1 of this paper, at a given temperature the eBW recipe 
significantly alters the feeddown of $\Delta$ to proton, neutron and pion yields, 
as well as the yield of $\Lambda$, compared to the Breit-Wigner scenario 
with fixed widths.
As the eBW approach was not applied to the system of pp collisions at 
$\sqrt{s} = 17.3$ GeV, such a test appears an attractive additional motivation for our analysis.

The paper is organized as follows.
The set of common particle yields of pp collisions at $\sqrt{s}$ = 17.3 GeV,
basing on the published results of NA49 and NA61/SHINE experiments, is proposed in Sec. 2. 
The main features of the statistical hadronization model used in our analysis
are described in Sec. 3.
The results of the model fits to the hadron yields are presented in Sec. 4, 
showing the influence of several calculation methods.
The summary in Sec. 5 closes the paper.

%---------------------------------------------
%-----------  experimental yields ------------
%---------------------------------------------

\section{Experimental particle multiplicities from NA49 and NA61/SHINE}

The NA49 and NA61/SHINE collaborations performed the extensive studies
of the inelastic proton-proton interactions at $\sqrt{s}$ = 17.3 GeV 
(laboratory beam momentum 158 GeV/c), acquiring, 
respectively, $4.8 \times 10^6$ and $5.3 \times 10^7$ events.
These fixed-target experiments were carried out using the beams from 
Super Proton Synchrotron (SPS) at CERN.
The previous experiments on proton-proton interactions in this energy range 
(starting from bubble chamber measurements) and their results were carefully 
reviewed by the NA49 collaboration \cite{NA49Alt2006}.
The experiments provided numerous observables, in between the 
rapidity- and $p_T$-integrated global particle multiplicities. 
These values included the feed-down from strong and electromagnetic decays
of other hadrons, 
but the contributions from weak decays were subtracted 
in these analyses \cite{NA61many}.
The particle yields used already in numerous model calculations 
(in particular statistical hadronization codes) were, in certain analyses, 
based on preliminary results (conference contributions and/or Ph.D. Thesis).
Some of them were confirmed, when the full analysis was completed 
and final results were published, where minor changes could be observed.
The multiplicities published by NA49 and NA61/SHINE collaborations are collected 
in table \ref{Tab:particle_yields}.
Only the values reported in the collaboration papers are included
in this table, except a few cases \cite{NA49Sikler1999, NA49Barton2001}.
The conclusion is that both data sets are mutually consistent within very 
few percent and well within the declared global uncertainties. 
Below we review these results, and show
in table \ref{Tab:common_yields}
the evaluated yields in case of data measured in both experiments,
while simply copying there
several yields obtained by one experiment only.

\begin{table}[t]
    \centering
    \begin{tabular}{|l|*{2}{c|l|}}
    \hline 
    Experiment & \multicolumn{2}{c|}{NA49} & \multicolumn{2}{c|}{NA61/SHINE}\\
       \hline
         Particle & Yield & Ref. & Yield & Ref. \\
         \hline
    $\pi^+$ &  $3.018$ & \cite{NA49Alt2006} &
               $3.110 \pm 0.030 \pm 0.26$ & \cite{NA61many} \\
    \hline
    $\pi^-$ &  $2.360$ & \cite{NA49Alt2006} &
               $2.404 \pm 0.034 \pm 0.18$ & \cite{NA61many} \\
    \hline 
    $K^+$ &  $0.2267 $ & \cite{NA49Anticic2010a} &
               $0.234 \pm 0.014 \pm 0.017$ & \cite{NA61many}\\
    \hline
    $K^-$ &  $0.1303 $ & \cite{NA49Anticic2010a} &
               $0.132 \pm 0.011 \pm 0.009$ & \cite{NA61many} \\
    \hline 
     $K^0_S$ &  $0.18 \pm 0.04$ & \cite{NA49Sikler1999} &   &   \\   
    \hline
     $K^\star (892)^0$ & $ (74.1 \pm 1.5 \pm 6.7)10^{-3}$ & \cite{NA49Anticic2011} & 
                        $ (78.44 \pm 0.38 \pm 6.0)10^{-3}$ & \cite{NA61Kstar} \\
    \hline
    $\overline{K^\star (892)}^0$ & $ (52.3 \pm 1.0 \pm 4.7)10^{-3}$ & \cite{NA49Anticic2011} & 
                         & \\
    \hline
    $\phi$ & $(12.0 \pm 1.5)10^{-3}$ &\cite{NA49Afanasiev2000} &
             $(12.56 \pm 0.33 \pm 0.32)10^{-3}$ & \cite{NA61phi} \\
    \hline 
    $p$   & $1.162 \pm 0.029$ & \cite{NA49Anticic2010b} & 
           $1.154 \pm 0.010 \pm 0.04$ & \cite{NA61many} \\
    \hline
    $\overline{p}$ &  $(38.6 \pm 1.3)10^{-3}$ & \cite{NA49Anticic2010b} &
               $ (40.2 \pm 2.0 \pm 3.0)10^{-3}$ & \cite{NA61many} \\
    \hline
    $n$ & $0.67 \pm 0.10$ &\cite{NA49Anticic2010b} & & \\
    \hline
     $\Lambda$ &  &  &
               $0.120 \pm 0.006 \pm 0.010$ & \cite{NA61lambda} \\
    \hline
     $\Lambda(1520)$ &  $(12 \pm 3) 10^{-3}$ & \cite{NA49Barton2001} &
                &  \\
    \hline
     $\Xi ^-$ &   &  & $ (3.3 \pm 0.1 \pm 0.6) 10^{-3}$ & \cite{NA61Ksi} \\
    \hline 
     $\overline{\Xi} ^+$ &  &  &  $(0.79 \pm 0.02 \pm 0.10)10^{-3}$ & \cite{NA61Ksi} \\
    \hline 
    \end{tabular}
    \caption{Mean total hadron multiplicities from NA49 and NA61/SHINE experiments. 
    The results obtained by the NA49 collaboration for abundantly produced light charged mesons are listed without the systematic uncertainties, as they 
    were provided for spectral analysis only (see text), and statistical accuracy was found to be negligible. 
    The total yield of neutrons is extracted from results of \cite{NA49Anticic2010b}, as described in the Appendix A. }
    \label{Tab:particle_yields}
\end{table}

While the NA61/SHINE results are always described with statistical and systematic 
uncertainties, the NA49 Collaboration provided a single value in most cases.
Therefore some comments are necessary.
In the case of copiously produced light charged mesons 
($\pi^\pm$ and K$^\pm$),
the statistical uncertainties of the differential cross sections
measured by NA49 were very small, 
and their relative systematic uncertainties were determined to be 2\% - 2.2\%.
They were treated as the relative uncertainties of total yields for 
those particles in numerous data sets.
However, as noticed already in \cite{NA61Kstar}, 
systematic uncertainties of the total yields were not reported 
in the NA49 papers.
The differences between the NA49 and NA61/SHINE results 
for light charged mesons are often within 
the statistical accuracy of the measurements.
As the systematic uncertainties of the total yields obtained 
by NA61/SHINE were carefully evaluated \cite{NA61many}, 
we propose to use the latter values with their uncertainty estimates 
for the light charged mesons.
The multiplicity of $\pi^-$ equal to 2.444 $\pm$ 0.130 \cite{NA61hminus}
obtained by the NA61/SHINE collaboration using the $h^-$ method 
(without particle identification) agrees with the result of 
the subsequent analysis \cite{NA61many} within 2 standard deviations.

In the case of yields provided by both NA49 and NA61/SHINE collaborations,
we averaged the two results, taking the inverse squared uncertainty of each 
measurement as weight.
The precision of NA61/SHINE measurement of the $\phi$ meson 
is much better compared to the NA49 result, 
so the former dominates the result of the average.
For the $\Xi$ and $\overline{\Xi}$ baryons the NA49 collaboration provided the 
results at midrapidity \cite{NA49Susa2002}, but not the global yield. 
In some of the previous works applying the thermal approach to the considered
system, the yields of $\Omega$ and $\overline{\Omega}$ were included to the 
analyses \cite{Becattini2006, Kraus2010, Begun2018}. 
However, since these yields were only preliminary \cite{BarnaPhd2002} and not 
published later, we excluded them from our data set.

The NA49 and NA61/SHINE are the spectrometers designed for charged particles.
The measurement of neutrons was realized in the NA49 set-up supplemented with the electromagnetic and hadronic calorimeter RCal, placed off the beam axis.
The distribution of the Feynman variable $x_F$
of neutrons was measured \cite{NA49Anticic2010b}. 
As the total yield was not published, we evaluate it in the Appendix A.
The NA61/SHINE setup is equipped with forward calorimeter PSD \cite{NA61detector}, 
but this detector was not yet installed, 
when the proton-proton measurements were performed.
The evaluated values of particle yields in proton-proton interactions at $\sqrt{s}$ = 17.3 GeV are presented in table \ref{Tab:common_yields}, 
which we propose to use as the common data set.

\begin{table}[t]
    \centering
    \begin{tabular}{|l|*{2}{c|l|}c|}
    \hline 
         Particle & Yield & Selection\\
    \hline
    $\pi^+$ &  $3.11 \pm 0.26$ & NA61/SHINE \\
    $\pi^-$ &  $2.40 \pm 0.18$ & NA61/SHINE \\
    $K^+$ &  $0.234 \pm 0.022$ & NA61/SHINE \\
    $K^-$ &  $0.132 \pm 0.014$ & NA61/SHINE \\
    $K^0_S$ &  $0.18 \pm 0.04$ & NA49 \\   
    $K^\star (892)^0$ & $(76.5 \pm 4.5)10^{-3}$  & average\\
    $\overline{K^\star (892)}^0$ & $(52.3 \pm 4.8)10^{-3}$ & NA49 \\
    $\phi$ & $ (12.5 \pm 0.4)10^{-3}$ & average\\
    $p$   & $1.159 \pm 0.024$ & average\\
    $\overline{p}$ &  $(38.8 \pm 1.2)10^{-3}$ & average \\
    $n$ & $0.67 \pm 0.10$ & NA49, see Appendix A\\
    $\Lambda$ & $0.120 \pm 0.012 $ & NA61/SHINE \\
    $\Lambda(1520)$ &  $(12 \pm 3)10^{-3}$ & NA49  \\
    $\Xi ^-$ & $(3.3 \pm 0.6)10^{-3}$ & NA61/SHINE\\
    $\overline{\Xi}^+$ &  $(0.79 \pm 0.10)10^{-3}$ & NA61/SHINE \\
    \hline
    \end{tabular}
    \caption{Evaluated set of multiplicities of particles emitted
    from proton-proton interactions at $\sqrt{s}$ = 17.3 GeV,
    based on the NA49 and/or NA61/SHINE results, 
    as indicated in the last column. }
    \label{Tab:common_yields}
\end{table}

\begin{table}[ht]
    \centering
    \begin{tabular}{|l|c|c|}
    \hline
         &  Initial & Reconstructed \\
         \hline
    Charge        & 2 & $1.93 \pm 0.32$ \\
    Baryon number & 2 & $1.92 \pm 0.07$ \\
    Strangeness   & 0 & $-0.011 \pm 0.027 $ \\
    \hline
    \end{tabular}
    \caption{Conserved quantities in proton-proton interactions evaluated from 
    the yields listed in table \ref{Tab:common_yields} }
    \label{tab:conservation}
\end{table}

In table \ref{tab:conservation} we summarize the values of total charge, 
baryon number and strangeness built from the yields of measured particles 
as listed in table \ref{Tab:common_yields}. 
The dominant contribution to the uncertainty of charge is due to the pion yields. 
A comparison with the initial values of Q, B and S shows, 
that the collected data set clearly covers most of the produced matter. 
The influence of heavier mesons and baryons that are not listed 
would not contribute significantly due to low expected multiplicities.
The small gap in charge and baryon number might be bridged with 
deuterons (with multiplicity around a few percent).
The statistical model estimation of their yield, 
presented further in this paper, supports this conjecture.
Deuterons are clearly seen in the particle identification plot of NA61/SHINE TPC detector at momenta  below 2 GeV/c (cf. figure 4 in \cite{NA61many}).

%---------------------------------------------
%-----------  model ----- --------------------
%---------------------------------------------

\section{Statistical hadron gas model}

An ample literature is devoted to the statistical treatment of hadron emission from subatomic collisions (see e.g. \cite{Hagedorn85,Cleymans91,Letessier95,Becattini98,Redlich02,Andronic06,Becattini2006,Mekjian07,Kraus07,Vovchenko18}). 
Here we shall sketch the approaches considered for our analysis. We apply the model of Hadron Resonance Gas without excluded volume corrections, kindly provided by the Thermal-FIST package \cite{ThermalFIST}. The yields of baryons and antibaryons are expected to be small, and these involving (anti-)strange quarks even smaller. Therefore in our calculations we exclude the full Grand Canonical ensemble, and consider two approaches: (i) Grand Canonical ensemble for all the hadrons except 
for the strangeness sector, where the Canonical treatment would enforce the strict conservation of $S$, which due to two protons in the initial channel amounts to 0. This approach shall be named "SC". (ii) Canonical ensemble including all the hadrons. This approach ensures in addition the exact conservation of baryon number and charge, which due to the initial channel amount to $B$ = 2 and $Q$ = 2. For this approach we shall use the "CE" symbol throughout the paper. Also, an additional $\gamma_s$ parameter is introduced that accounts for possible non-equilibration of hadrons containing strange quarks: if $\gamma_s < 1$, their yields are suppressed compared to the equilibrium values \cite{Rafelski91,Letessier95}.

Within the Grand Canonical ensemble for the system of hadrons inside volume $V$ and in temperature $T$, the partition function modified to account for the possible undersaturation of strangeness can be represented as follows:

\begin{equation}
\ln Z^{GC} (T, V, \vec{\mu}, \gamma_S) 
= \sum_j Z_j^{1,GC} (T, V, \vec{\mu}, \gamma_S)
\end{equation}

\noindent where $\vec \mu = (\mu_S, \mu_B, \mu_Q)$ is the vector of chemical potentials related to the total strangeness, baryon number, and charge of the system\footnote{the charm production is neglected in our analysis}, whereas the $j$ index iterates over all the hadrons. $Z_j^{1,GC}$ is the one-particle partition function, which for the $j$-th hadron with energy $E_j$ reads:

\begin{equation}
Z_j^{1,GC} = \pm \frac{V g_j}{2\pi} 
\int_0^\infty p^2 dp \int \rho_j (m) dm
 \cdot \ln \left[ 
    1 \pm \gamma_s^{|s_j|}
    \exp \left( \frac{\vec q_j \cdot \vec \mu_j - E_j }{T}\right)
  \right]
  \label{eq:Z1GC}
\end{equation}

\noindent where the +(-) sign is for a fermion (boson), $g_j$ is the hadron's spin degeneracy, $\vec q = (s_j, b_j, q_j)$ combines, respectively, the strangeness and baryon numbers as well as charge of the hadron, whereas $s_j$ is the number of strange plus anti-strange valence quarks contained in this hadron. Finally, $\rho_j (m) = \delta(m - m_j)$ unless the hadron in question is characterized by the mass distribution. This issue will be discussed in the further part of the Section. 

An usual approach to obtain the partition function describing the canonical ensemble with respect to given conserved charges is the projection onto a subsample of microstates that keep these charges strictly \cite{Hagedorn85,Cleymans91,Cleymans99}.
One can define the vector of generalized charges $\vec Q = (S, B, Q)$ and thus say that for the CE approach all its elements are constrained, whereas within the SC method only the first one is fixed, and the rest is conserved only on average. In such a representation a concise formulation of the partition functions describing both approaches is possible:

\begin{equation}
  Z^{CE/SC} (\vec Q) = \prod_{n = 1}^{n_{max}} \left[ 
    \frac{1}{2\pi}  \int_{-\pi}^{\pi} d\phi_n e^{ -i Q_n \phi_n }
  \right]
  \circ 
  \exp \left( \sum_j Z_j^{1,CE/SC} \right)
\end{equation}

\noindent where $\vec \phi = (\phi_S, \phi_B, \phi_Q)$ are the auxiliary angles of the projection operation, and $n$ iterates over the conserved charges up to $n_{max}$ = 1 for the SC and 3 for the CE approach. The one-particle partition function $Z^{CE/SC} (\vec Q)$ can be obtained by the following modification of Eq.~\ref{eq:Z1GC}: the fugacity terms related to the $n$-th conserved charges ($\exp ^{~} (\mu_n /T)$ terms) have to be replaced by $\exp ^{~} (i \phi_n)$. In addition, for systems where the overall number of strange hadrons is small, the strangeness could occupy some volume $V_\mathrm{C}$, which is in principle different than $V$. This modification can be introduced for hadrons with $s_j \neq 0$ at the level of their one-particle partition functions.
It is often represented in terms of radii $R$ and $R_\mathrm{C}$
of spheres of the respective volumes.

For the calculation of yields from the p-p collisions at $\sqrt{s} = 17.3$~GeV the list of hadrons included also the light nuclei up to $^4$He and their respective hypernuclei. We also excluded hadrons containing charm quarks, based on finding that their influence on the fitted thermal parameters is negligible. For the CE approach we applied the quantum statistics only to mesons, letting the momentum distributions of baryons be described by the Boltzmann function. We have checked, that the impact of this approximation is negligible. The experimental hadron yields delivered by the NA49 and NA61/SHINE Collaborations are given after including the strong and
electromagnetic decays of other particles, but before the weak decays. 
The model approach follows the same convention.

A default parametrization of the mass distribution of a resonance is the relativistic Breit-Wigner (BW) function:

\begin{equation}
  \rho_j(m) = A_j 
    \frac{2 m \, m_j \, \Gamma_j}
         {(m^2 - m_j^2)^2 + (m_j \, \Gamma_j)^2}
\end{equation}

%\begin{equation}
%  \left. \frac{dN}{dm} \right|_{R \rightarrow ab} ~=~ A
%    \frac{2 m ~ m_0 ~ \Gamma_{R \rightarrow ab} (m)}
%         {(m^2 - m_0^2)^2 ~+~ (m_0 {}^{~} \Gamma_R)^2}
%\end{equation}

\noindent where $A_j$ is the normalization constant, $m_j$ is the centroid, and $\Gamma_j$ -- the decay width. However, if the $j$-th hadron has $k$ different decay channels, each of them has its own threshold mass, $m^{\mathrm{thr}}_{j \rightarrow k}$. 
The problem of choice of the lower limit of integration can be to some extent accounted for by weighting all the $m^{\mathrm{thr}}_{j \rightarrow k}$ with the respective branching ratios, $b_{j \rightarrow k}$.
This recipe is implemented in \cite{ThermalFIST} as the "energy-independent BW approach". 
However, it does not account for the fact that the momentum space available to the
products of decay in channel $k$ increases with the excess of substrate's mass 
over the threshold value. 
It is possible to include both these effects by introducing the energy-dependent partial widths for each decay channel $\Gamma_{j \rightarrow k} \sim (q^2)^{L_{j \rightarrow k} + \frac{1}{2}}$ [Jackson64,PDG], where $q$ is the momentum difference between decay products in the substrate's reference frame (positive only above the threshold), and $L_{j \rightarrow k}$ is the angular momentum released in that decay. This recipe was approximated in \cite{SHARE} and further reformulated in \cite{Vovchenko18}, such that $\Gamma_{j \rightarrow k}$ from the momentum-dependent function became mass-dependent only, 
and equal to the PDG value of the total width times the branching ratio 
($\Gamma_j^\mathrm{pdg} \cdot b^{\mathrm{thr}}_{j \rightarrow k}$)
at the centroid of the BW distribution ($m = m_j^\mathrm{pdg}$): 

\begin{equation}
 \Gamma_{j \rightarrow k} (m) = 
 \Gamma_j^{\mathrm{pdg}} b^{\mathrm{pdg}}_{j \rightarrow k} \cdot
 \left[ 
 \frac{1 - \left( \frac{m^{\mathrm{thr}}_{j \rightarrow k}}{m}        \right)^2}
      {1 - \left( \frac{m^{\mathrm{thr}}_{j \rightarrow k}}{m_j^{\mathrm{pdg}}} \right)^2} 
 \right]
        ^{L_{j \rightarrow k} + \frac{1}{2}}
\end{equation}

%\begin{equation}
% \Gamma_{R \rightarrow ab~} (m) ~=~ 
% \Gamma_R^{\mathrm{pdg}} ~\cdot~ BR^{\mathrm{pdg}}_{R \rightarrow ab} ~\cdot~
% \left[ 
% \frac{1 - \left( \frac{m^{\mathrm{thr}}_{a+b}}{m}        \right)^2}
%      {1 - \left( \frac{m^{\mathrm{thr}}_{a+b}}{m_R^{\mathrm{pdg}}} \right)^2} 
% \right]
%        ^{L_{R \rightarrow ab} ~+~ \frac{1}{2}}
%\end{equation}

\noindent This approach is named the "energy-dependent BW scheme", and was reported to significantly change the values of thermal parameters extracted from at least some of the experimental yields \cite{Vovchenko18}. 
Since to our knowledge this improvement was not applied to the case of p-p collisions at $\sqrt{s} = 17.3$ GeV, it becomes interesting to check its effect 
on these data.

%---------------------------------------------
%-----------  results --- --------------------
%---------------------------------------------

\section{Statistical hadronization results}

While the grand canonical model of statistical hadronization might be applied 
to the collisions of heavy nuclei (Au+Au or Pb+Pb), 
the small volume of the proton-proton system calls for limitations: 
application of the canonical model 
only for strange particles or for all hadrons.
In the analysis we apply these two approaches:
(i) grand canonical ensemble for non-strange particles 
and canonical one for strange particles (SC), and
(ii) the canonical ensemble (CE).
Both models appear in two versions: 
standard treatment of the resonance shape (BW), 
so-called the energy-independent Breit-Wigner shape, 
and the improved version of the energy-dependent 
Breit-Wigner profile (eBW), as described in the previous chapter.
In our analysis we take into account all the measured particles, 
as listed in table \ref{Tab:common_yields}.
Following the previous studies \cite{Kraus2010,Begun2018}, 
we consider also the option of removing the $\phi$-meson yield, 
although the production of this meson was measured independently 
and the results are in good agreement.

\begin{table}[htb]
    \centering
    \begin{tabular}{*{7}c}
       \hline
Data & Option & T (MeV) & $\mu_B$ (MeV) & R (fm) & $\gamma_S$ & $\chi^2/NDF$\\
\hline
\multicolumn{7}{c}{Grand Canonical - Strangeness Canonical (SC)} \\
\hline
all & BW & 155.9 $\pm$ 2.3 & 231.1$\pm$2.9 & 1.78 $\pm$ 0.08 & 0.45$\pm$0.01 & 145/11=13 \\
%$\frac{145}{11}=13$ \\
    & eBW & 165.1 $\pm$ 3.0 & 242.5$\pm$3.4 & 1.59 $\pm$ 0.09 & 0.42$\pm$0.01 & 120/11=11 \\
    % $\frac{120}{11}=11$ \\
         \hline
no $\phi$ & BW  & 168.7 $\pm$ 3.0 & 240.3$\pm$3.5 & 1.35 $\pm$ 0.08 & 0.65$\pm$0.02 & 12.5/10=1.3  \\
%$\frac{12}{10}=1.2$ \\
        & eBW  & 176.1 $\pm$ 3.5 & 250.5$\pm$3.9 & 1.29 $\pm$ 0.08 & 0.57$\pm$0.02 & 11.0/10=1.1 \\
        %$\frac{11}{10}=1.1$\\
\hline
\multicolumn{7}{c}{Canonical (CE)} \\
\hline
 all & BW & 163.4 $\pm$ 2.8 & & 1.64 $\pm$ 0.09 & 0.42$\pm$0.01 &
 143/12=12 \\ %$\frac{143}{12}=12$ \\
     & eBW & 174.5 $\pm$ 3.8 & & 1.44 $\pm$ 0.09 & 0.39$\pm$0.01 & 117/12=9.7 \\ %$\frac{117}{12}=9.7$ \\
         \hline
no $\phi$ & BW  & 176.0 $\pm$ 3.3 & & 1.29 $\pm$ 0.07 & 0.59$\pm$0.02 & 16.0/11=1.5 \\ %$\frac{16}{11}=1.4$ \\
       & eBW  & 184.7 $\pm$ 3.9 & & 1.20 $\pm$ 0.08 & 0.52$\pm$0.01 & 15.8/11=1.4 \\ %$\frac{16}{11}=1.4$ \\
\hline
    \end{tabular}
    \caption{Hadron-resonance gas model parameters for proton-proton collisions at $\sqrt{s} = 17.3$ GeV evaluated in the SC and CE versions of the statistical hadronization code Thermal-Fist \cite{ThermalFIST}. }
    \label{tab:SCplusCE}
\end{table}

The results for the SC version of the statistical hadronization model 
are summarized in table \ref{tab:SCplusCE}.
We find that for the data listed in table \ref{Tab:common_yields}  
the $\chi^2/NDF$ of the fit reaches unacceptable values, 
whereas the exclusion of the $\phi$ meson yield 
results in drastic improvement of the fit quality, to the reasonable level, as observed previously.
The application of the eBW approach in the treatment 
of broad states close to the threshold mass in the exit channel
does not improve the situation significantly, 
although the trend is positive.
The use of eBW has the following consequences: 
(i) the extracted temperature is raised by around 8 MeV, 
(ii) the baryochemical potential is raised by about 10 MeV, 
(iii) the radius R is by $\sim$0.13 fm smaller, and 
(iv) the strangeness suppression factor is reduced by $\sim$0.05.
The most similar analysis to which these results can be compared 
are the fits of the SC model to two sets of NA49 data, 
named A2 and B2 in \cite{Kraus2010}, as specified in table 4 therein.
A comparison of those fits to our analysis within the BW parameterization
and the "no $\phi$" scenario exhibits mutual agreement within 3 standard deviations.

Compared to the SC model, which explicitly requires only the strangeness balance,
the canonical ensemble requires also the conservation of other conserved quantities.
The results of the Thermal-Fist model calculations in the CE approach, 
shown in the lower rows of table \ref{tab:SCplusCE},
provide the following conclusions: 
on average the extracted temperatures are raised by about 8 MeV,
the radii are dropped by about 0.1 fm, and $\gamma_s$ are decreased slightly by about 0.04.
The exclusion of the $\phi$ meson leads to the same effect as observed previously,
namely the $\chi^2 / NDF$ values fall from unacceptable to reasonable value.

Our findings in the BW version of the CE approach can be compared to 
the results obtained in \cite{Begun2018}, 
separately for NA49 and NA61/SHINE (respectively, 
Tables III and IV therein).
Overall the results are similar. 
However, whereas the values for the NA61/SHINE experimental yields
are consistent with our findings within 2 standard deviations,
the case for NA49 is often outside the 3 sigma range.
Also, within the "no $\phi$" scenario, for which the quality of fits 
is reasonable, the calculations for the data set 
proposed in our analysis result in better $\chi^2 / NDF$ values. 
Taking into account the mass-dependent BW scheme additionally used
in our analysis, we may conclude that some improvement in data, model and fit
quality was achieved in the current approach, 
although the $\phi$ meson puzzle 
is not resolved yet at this stage.

\begin{table}[ht]
    \centering
    \begin{tabular}{*{6}c}
    \hline
     T (MeV) & $\mu_B$ (MeV) & R (fm) & R$_C$ (fm) & $\gamma_S$ & $\chi^2/NDF$\\
     \hline
173.7 $\pm$ 3.2 & 248.5$\pm$3.7 & 1.35 $\pm$ 0.08 & 1.82 $\pm$ 0.13 & 0.41$\pm$0.01 & 11.5/10=1.15 \\
     \hline
    \end{tabular}
    \caption{Hadron-resonance gas parameters 
     for proton-proton collisions at $\sqrt{s} = 17.3$ GeV
     in the SC version of the statistical hadronization model, 
     when $R_C$ radius was added to the fit parameters.
     The eBW approach was used.
     All the particle yields listed in table \ref{Tab:common_yields}
     are taken into account. }
    \label{tab:hadrongasRR}
\end{table}

Searching for alternatives to the unphysical removal of 
the well established experimental $\phi$ meson yield,
we explored the possible consequences of different volume occupied by 
open strange hadrons than that for all other particles. 
Here we remind the naming of radii of spheres of those volumes: 
$R_C$ for open strangeness, and $R$ for the non-strange hadrons.
We apply only the eBW ansatz, treating it as an improvement of the model description.
In the SC approach where the $R_C$ was set as free parameter in Thermal-FIST, 
we observe the very significant improvement 
of the global quality of the fit: 
the $\chi^2$ is close to the number of degrees of freedom (NDF), 
as quoted in table \ref{tab:hadrongasRR}.
The $\chi^2$ is reduced by one order of magnitude with respect to the 
SC model calculations implying $R = R_C$ and all the measured yields,
including the $\phi$-meson.

\begin{table}[ht]
    \centering
    \begin{tabular}{ccr}
    \hline 
         Particle &  Yield & $\frac{yield-exp}{\sigma}$\\
    \hline
    $\pi^+$ &  $3.34$ & $0.90$ \\
    $\pi^-$ &  $2.44$ & $0.18$ \\
    $K^+$ &  $0.239$ & $0.24$ \\
    $K^-$ &  $0.118$ & $-1.01$ \\
    $K^0_S$ &  $0.21$ & $0.77$ \\   
    $K^\star (892)^0$ & $7.87 \times 10^{-2}$  & $0.47$\\
    $\overline{K^\star}(892)^0$ & $4.35 \times 10^{-2}$ & $-1.84$\\
    $\phi$ & $ 1.25 \times 10^{-2}$ & $0.00$\\
    $p$   & $1.143$ & $-0.69$\\
    $\overline{p}$ &  $3.89 \times 10^{-2}$ & $0.09$ \\
    $n$ & $0.87$ & $1.98$\\
    $\Lambda$ & $0.135$ & $1.23$ \\
    $\Lambda(1520)$ &  $1.13 \times 10^{-2}$ & $-0.22$  \\
    $\Xi ^-$ & $3.24 \times 10^{-3}$ & $0.04$\\
    $\overline{\Xi}^+$ &  $7.9 \times 10^{-4}$ & $0.00$ \\
    \hline
     $\pi^0$ &  $3.16$ & \\
     $\eta$ &   $0.296$ & \\
     $\rho^0(770)$ & $0.315$ \\
     \hline
     $D^0$ &  $1.72 \times 10^{-3}$ &\\
     $\overline{D}^0$ & $1.45 \times 10^{-3}$ & \\
     $D^+$ & $8.62 \times 10^{-4}$ &\\
     $D^-$ & $5.51 \times 10^{-4}$ &\\
     $D_S$ & $3.71 \times 10^{-4}$ &\\
     $J/\psi$ & $2 \times 10^{-6}$ &\\
     $\psi(2S)$ & $8.5 \times 10^{-8}$ &\\
     \hline
     $\overline{n}$ & $5.08 \times 10^{-2}$ & \\
     $\overline{\Lambda}$ & $1.27 \times 10^{-2}$ &\\
     $\Sigma(1385)^+$ & $2.05 \times 10^{-2}$ &\\
     $\overline{\Lambda}(1520)$ & $1.09 \times 10^{-3}$ & \\
     $\Xi(1530)$ & $1.54 \times 10^{-3}$ &\\
     $\overline{\Xi}(1530)$ & $2.55 \times 10^{-4}$ &\\
     $\Omega$ & $7.6 \times 10^{-5}$ & \\
     $\overline{\Omega}$ & $3.8 \times 10^{-5}$ & \\
     $d$ & $2.58 \times 10^{-2}$ & \\
     $\overline{d}$ & $5.00 \times 10^{-5}$ & \\
     $\Lambda_c^+$ & $2.43 \times 10^{-5}$ & \\
     \hline
    \end{tabular}
    \caption{Particle yields evaluated according to the SC approach 
    of the statistical model, with energy-dependent Breit-Wigner shape,
    for proton-proton interactions at $\sqrt{s}$ = 17.3 GeV.
    The model parameters are taken from table \ref{tab:hadrongasRR}.
     In the upper part of the table, the right column shows 
    the difference to the experimental values in table \ref{Tab:common_yields},
    evaluated relative to the experimental uncertainty.}
    \label{tab:final}
\end{table}

The same attempt of freeing the $R_C$ parameter was not successful 
in the case of the canonical formulation of statistical hadronization, 
as the $\chi^2$ values remained at approximately the same, 
very high level.
In this procedure both volume parameters, $R$ and $R_C$, 
were found to be in close agreement 
(within the accuracy of the fit), and all the fit parameters
were compatible within 1 standard deviation 
to those obtained in the CE scenario where $R_C = R$ was enforced. 
Therefore, we skip this approach from our considerations.

Assuming the correctness of the model, 
the canonical radius for strange particles $R_C$ obtained in the fit 
is found to be around 0.5 fm larger than the $R$ parameter corresponding to 
the non-strange particles. 
The $R_C$ radius being significantly larger than $R$ might be related to 
the larger effective volume of the emission of strange particles,
which can be measured via the femtoscopy of strange mesons.
Many measurements of the HBT interferometry for pp system
were performed (for the systematics, see \cite{Chajecki09}), 
but the particles used  were mainly charged pions.
While kaons are quite commonly used to establish the source size in nucleus-nucleus collisions,
only two measurements report the kaon interferometry results from pp interactions.
In the measurement at $\sqrt{s} = 27.4$ GeV \cite{LEBC1992} 
the gaussian radius of the $K^\pm$ source equals to $1.87 \pm 0.33$~fm
(in the Kopylov-Podgoretsky parametrization \cite{KP}), 
compared to $1.71\pm0.04$~fm obtained for pions.
The uncertainty of the former is quite substantial, 
so the measurement provides an indication only,
that it is larger than pion source radius.
Similarly, the CMS collaboration  at $\sqrt{s} = $~900 GeV \cite{Dogra14} 
obtained the kaon emitting source well above 2 fm, 
while the radius obtained for pions is smaller and equal to about 1.5 fm 
(here the difference is above the uncertainty range).
Therefore the larger value of $R_C$ obtained in the actual fit to the 
particle yields can be viewed as in-line with the available HBT measurements.

Overall, for the current data set three scenarios of the eBW version
of the hadron-resonance gas model deliver the acceptable values of 
$\chi^2 / NDF$: 
one SC variant where $R_C$ is subject to free fitting,
and two approaches (SC and CE) where $R_C = R$ was enforced, 
but $\phi$ mesons were excluded. 
Assuming the overall correctness of the model,
we find the temperature parameter to be roughly within 174--185 MeV, $\mu_B$ 
staying around 250 MeV, the $R$ radius around 1.20--1.35 fm,
and the $\gamma_S$ parameter in-between 0.41--0.57.

Using the best-fit parameters of the statistical hadronization model
(values for the SC approach reported in table \ref{tab:hadrongasRR}), 
we have evaluated the particle yields and placed them in Table \ref{tab:final}.
Compared to the experimental values in table \ref{Tab:common_yields}, 
the differences never exceed twice the uncertainty of the measurement 
(the last column of the table \ref{tab:final}), whereas in most cases the
calculated yield is well within one standard deviation.
The principal deviation is observed for neutrons. 
The predicted yields for some not yet measured particles are also provided.

%---------------------------------------------
%-----------  conclusions --------------------
%---------------------------------------------

\section{Conclusions}

The common data set of particle yields produced 
in proton-proton collisions at $\sqrt{s}$ = 17.3 GeV
has been evaluated on the basis of complementary measurements 
of NA49 and NA61/SHINE collaborations.
This data set was used in statistical hadronization code 
Thermal-Fist \cite{ThermalFIST}.
We confirm the unacceptably high value of $\chi^2$/NDF, 
if all the particle yields are accounted for, both in the canonical 
and grand canonical - strangeness canonical formulation 
of the statistical hadronization model, 
and within assumption that each type of hadrons 
occupies the same volume.
The use of the modified shape Breit-Wigner parametrization for
the mass distributions of unstable particles, 
which might be important for broad resonances close to their 
decay mass threshold, 
slightly improves the description of particle yields in 
all the model formulations, but clearly does not solve the problem of 
unsatisfactory quality of the model fit to the experimental data.
We confirm, what was observed by several authors, 
that exclusion of the $\phi$ meson yield improves 
the fit quality to the $\chi^2/NDF$ value in the range of $1.1 - 1.5$.
However, the experimental yields of the $\phi$ meson, 
obtained in NA49 and NA61/SHINE measurements, 
are consistent and should not be unaccounted for.
The full experimental data set of particle yields
can be reasonably fitted if radial parameters $R$ and $R_C$ 
(describing the canonical volume of 
hadrons without and with open strangeness
in the statistical hadronization model) 
are allowed to vary independently.
Within this approach we find the canonical radius $R_C$ to be around
0.5 fm larger than the $R$ parameter.
This effect is compatible with the results of 
Hanbury-Brown-Twiss interferometry measurements 
performed for the pp system 
at $\sqrt{s} = $ 27.4 MeV by \cite{LEBC1992} 
and $\sqrt{s} = $ 900 MeV by \cite{Dogra14}.
In these measurements the invariant radius of the K-meson 
emitting source is above that obtained for pions.
The model with the best-fit parameters allows to predict 
the yields of several not yet measured particles
emitted from p+p at $\sqrt{s}$ = 17.3 GeV.

\bigskip
{\bf Acknowledgments}

We thank Katarzyna Grebieszkow for the valuable comments on the evaluation of the experimental data from NA49 and NA61/SHINE collaborations.

%\clearpage

\appendix
\section{Evaluation of the neutron multiplicity}

The NA49 collaboration made the measurement of the neutron yield \cite{NA49Anticic2010b}, 
however the integrated yield was not provided.
The $x_F$ distribution (figure 63 and table 11 in \cite{NA49Anticic2010b})
allows for an attempt to obtain this result. 
A few points in $x_F$ are not equidistant, which generates some 
flexibility in the selection of the bin size.
The details of our attempt are shown in table \ref{tab:neutron}. 

With the sample of $4.8 \times 10^6$ events the statistical errors are
negligible. 
It is reflected in the small inter-point differences in 
Figs. 7 and 63 of \cite{NA49Anticic2010b}.
The error bars in figure 63 therein are due to systematic uncertainties,
and table 2 lists their relative contributions. 
However, this list concerns the overall sample (not the data points), 
and no information can be drawn on correlations of these contributions.
Therefore, the error bars in figure 63 cannot be treated as independent,
and consequently, for the estimation of the uncertainty 
of the total yield, they cannot be added quadratically. 
In this situation we propose to add the error bars linearly, 
which results in 15\% overall systematic error, 
a value between the quadratic sum 
and the upper limit presented in table 2 of \cite{NA49Anticic2010b}.
Consequently, we obtain the neutron multiplicity equal to 
$0.67 \pm 0.10$ 
(the sum of values in the last column is multiplied by 2), 
very close to the value used already in \cite{Begun2018}.

\begin{table}[h]
    \centering
    \begin{tabular}{ccc||c|c}
    \hline
    $x_F$ & $\frac{dN}{dx_F}$ & $\Delta$ (\%) & Range & $\frac{dN}{dx_F}\Delta x$ \\
    \hline
    0.1	& 0.481	& 20.8 & $0.00-0.15$ &	$0.0722 \pm	0.0150$ \\
    0.2	& 0.407	& 14.7 & $0.15-0.25$ &	$0.0407	\pm 0.0060$ \\
    0.3	& 0.378	& 13.2 & $0.25-0.35$ &	$0.0378	\pm 0.0050$ \\
    0.4	& 0.325	& 11.5 & $0.35-0.45$ &	$0.0325	\pm 0.0037$ \\
    0.5	& 0.325	& 12.3 & $0.45-0.55$ &	$0.0325	\pm 0.0040$ \\
    0.6	& 0.293	& 10.2 & $0.55-0.65$ &	$0.0293	\pm 0.0030$ \\
    0.75& 0.286	& 10.5 & $0.65-0.85$ &	$0.0572	\pm 0.0060$ \\
    0.9	& 0.215	& 27.9 & $0.85-1.00$ &	$0.0323 \pm	0.0090$ \\
    \hline
    \multicolumn{4}{r}{sum}     &  $0.3344 \pm 0.0517$ \\
    \end{tabular}
    \caption{Evaluation of the neutron global yield in p+p interactions at $\sqrt{s}$ = 17.3 GeV.
    The data from the table 11 in \cite{NA49Anticic2010b} are listed in the three first columns.
    The bin range (column 4) was arbitrarily selected for the first and a few last bins.}
    \label{tab:neutron}
\end{table}

\end{document}